\documentclass[]{elsarticle}

\usepackage{lineno,hyperref}
\usepackage{siunitx}
\usepackage{booktabs}
\usepackage{tabularx}
\modulolinenumbers[5]

\journal{Astronomy and Computing}

\usepackage{framed}
\usepackage{listings}
\usepackage{color}
 
\begin{document}

\begin{frontmatter}

\title{An Empirical Evaluation On the Applicability of the DALiuGE Execution Framework}

\author[gzu]{Ying Mei}
\author[kmust]{Shoulin Wei}
\author[gzu]{Feng Wang\corref{mycorrespondingauthor}}
\cortext[mycorrespondingauthor]{Corresponding author}
\ead{fengwang@gzhu.edu.cn}
\author[icrar]{Chen Wu}
\author[icrar]{Rodrigo Tobar}
\author[pawsey]{Mohsin Shaikh}
\author[gzu]{Hui Deng}
\author[kmust]{Wei Dai}
\author[kmust]{Bo Liang}
\author[icrar]{Andreas Wicenec}

\address[gzu]{Center for Astrophysics, Guangzhou University, China}
\address[kmust]{Computer Technology Application Key Lab of Yunnan Province, Kunming University of Science and Technology, China}
\address[icrar]{ICRAR, The University of Western Australia, Australia}
\address[pawsey]{Pawsey supercomputing center, Australia}

\begin{abstract}
The Square Kilometre Array (SKA) project is an international cooperation project to build the largest radio telescope worldwide. Data processing is one of the biggest challenges of building the SKA telescope.
As a distributed execution framework, the Data Activated Liu Graph Engine (DALiuGE) was proposed to be one of the candidates for addressing the massive data of the SKA. 
DALiuGE has many distinctive features, but its actual ability to handle scientific data is still not evident. 
In this paper, we perform an objective evaluation of the usability of DALiuGE concerning the execution performance, developer workload, and implementation difficulty of porting the SAGECal to DALiuGE. The evaluation results showed that the DALiuGE enables fast integration of astronomical software, but there are significant differences in the efficiency of different parallel granularities. Even with the deep optimization of the program, there is still a gap between the current DALiuGE and the traditional MPI in execution performance. Therefore, we come to a preliminary conclusion that the DALiuGE has no performance advantage in batch processing of massive data, while it may be more suitable for application scenarios with more customized computational tasks, such as SKA science regional centers.
\end{abstract}


\end{frontmatter}

%\linenumbers

\section{Introduction} \label{sec:intro}
The Square Kilometre Array (SKA) project\citep{taylor2012square,hall2005square} is an international effort to build the world’s largest radio telescope proposed to be built in Australia and South Africa. As a radio interferometer with 197 mid-frequency dishes and up to 512 stations ( 131072 low-frequency antennas), the SKA brings huge challenges in both engineering and scientific research towards building and delivering a unique instrument.

The Science Data Processor (SDP) for the SKA is the essential parts of the SKA telescope that in charge of generating final scientific level data. 
An updated system sizing estimate revealed the computational loads for 13.6 PFLOPS for the SKA1-Low high-priority science objectives (HPSOs) and 11.5 PFLOPS for SKA1-Mid\citep{broekema2020commodity,dewdney2019ska1, broekema2015square, bonaldi2018square}. 

The Data Activated Liu Graph Engine (DALiuGE), a generic, data-driven execution framework, is developed for processing large astronomical datasets \citep{wu2017daliuge}. Drop is basic idea of the DALiuGE and is representation of applications and data, \texttt{AppDrop} and \texttt{DataDrop} respectively. The DALiuGE includes an interface for expressing complex data-reduction pipelines consisting of data sets, algorithmic components and a run-time implementation to execute such pipelines on distributed resources. The full SKA scale testing of the DALiuGE proved that the DALiuGE owns a series of distinctive advantages\citep{wu2017daliuge}. 

However, many system architecture experts urgently demand to test the DALiuGE in more detail in order to have a more reasonable and fair analysis of the applicability for the final SKA data processing. 
In recent years, novel parallel and distributed computing models are proposed, reflecting advances in new computational devices and environments such as optical interconnects, programmable logic arrays, networks of workstations, radio communications, mobile computing, DNA computing, quantum computing, sensor networks and so on. These new computing models may also become key technologies for future SKA data processing. Before finding the most suitable computing model for SKA, it is necessary to carefully analyze various frameworks and their respective advantages and disadvantages through comparative experiments. After all, we have never dealt with such a huge amount of data before. 

In this study, we mainly focus on a more objective analysis and evaluation of applicability of the DALiuGE for the SKA data processing. We strive to answer four key questions, including: 1) how about the programming difficulty on the DALiuGE? 2) is it easy to port existing and typical astronomical software? 3) how much is the porting cost? 4) what is the performance under the DALiuGE execution framework? In the rest of the paper, we first introduce SAGECal software and conduct static analysis and workload characterization of SAGECal in  Section 2. We propose three porting approaches in Section 3. The performance evaluation is presented in detail in Section 4. We finally discussed the relevant questions in Section 5 and came to a conclusion in the last Section. 

\section{A Brief Analysis of SAGECal}\label{sagel_analysis}

To objectively evaluate the DALiuGE, we initially consider to implement a test program for the DALiuGE. But we soon gave up this idea because we realized that this would complicate the comparison with other frameworks. Using a proven software and running it on different platforms would make usability evaluations more descriptive.

After long-term research and development, there are many reliable, mature and open sources radio data processing softwares such as CASA, AIPS/AIPS++, Miriad, SAGECal and so on. We finally chose SAGECal as our testing program. 

SAGECal is a self-calibration astronomical software that uses the Expectation Maximization (EM) algorithm to obtain the maximum likelihood estimation of the instrument and sky parameters \citep{yatawatta2009radio}. SAGECal uses an improved EM algorithm known as the SAGE algorithm\citep{fessler1994space} in order to speed up the convergence, reduce the computational complexity and improve the quality of calibration.

The reason we choose SAGECal is that it is a representative radio data processing software that has been widely used in data calibration. More importantly, a MPI based SAGECal (SAGECal-MPI) algorithm has been implemented \citep{yatawatta2015distributed}, which exploits data parallelism across the frequency axis. This would make it less difficult for us to port SAGECal-MPI to the DALiuGE.

\begin{figure}[ht!]
\centering
\includegraphics[scale=0.4]{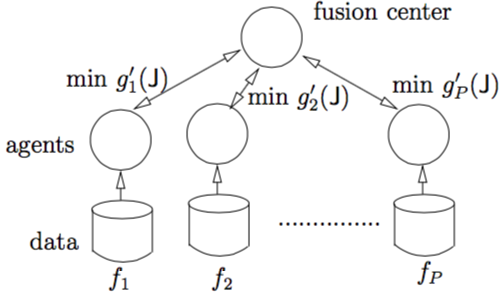}
\caption{The basic idea of SAGECal-MPI. Figure extracted from~\citep{yatawatta2015distributed}}
\label{fig:distributed_SAGECal}
\end{figure}

The central idea of SAGECal-MPI is illustrated in Figure \ref{fig:distributed_SAGECal}. Each agent is allocated a frequency, and works independently on minimizing a cost function \(g'_i(J)\) without interacting with others. During each iteration, the fusion center enforces the smoothness consensus from all agents by solving a consensus optimization problem by using the consensus alternating direction method of multipliers (C-ADMM). The distributed SAGECal algorithm is somewhat ``insensitive'' to the data size, working as effectively for large datasets as for small ones. By intentionally varying the total \(P\) number of agents, one can achieve an optimal balance between the completion time and resource footprint (i.e. degree of data parallelism).

\begin{figure}[ht!]
\centering
\includegraphics[scale=0.37]{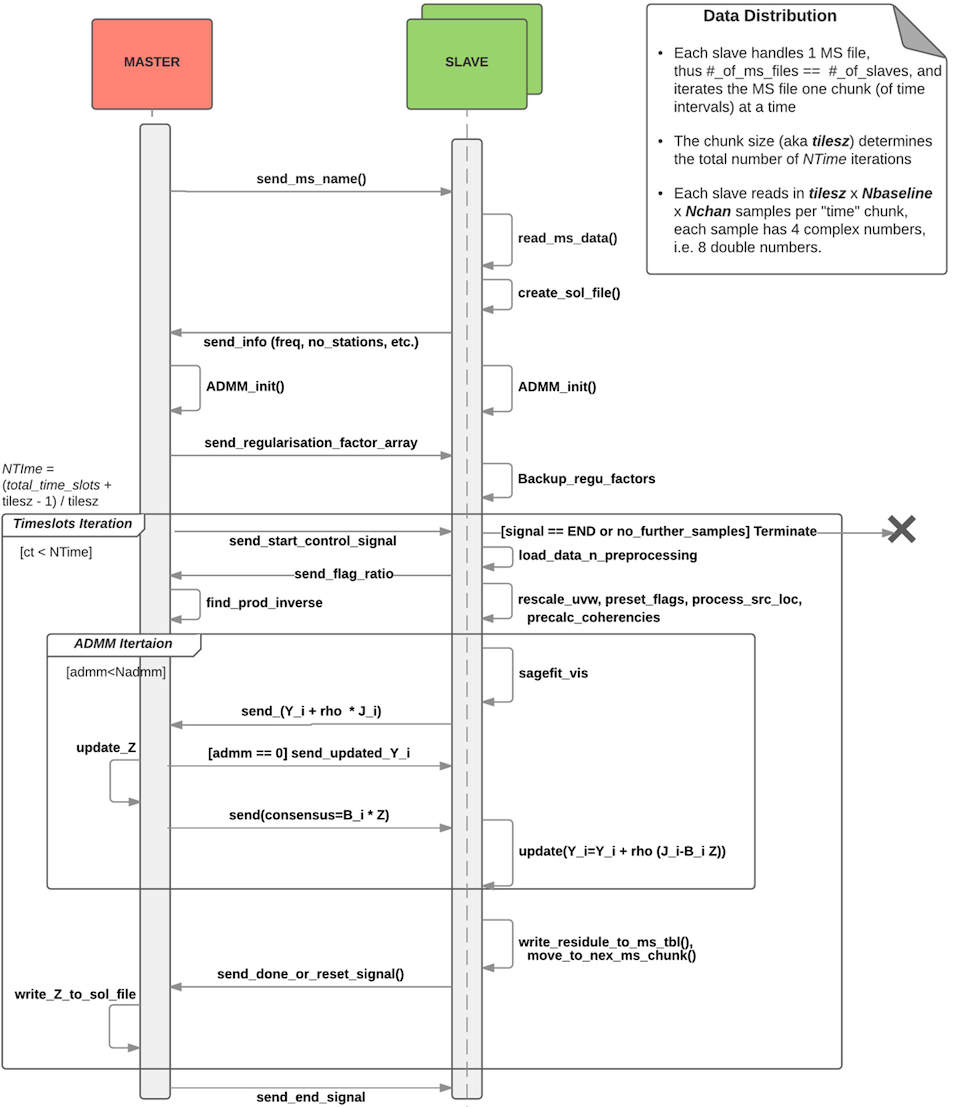}
\caption{A UML sequence diagram showing message exchanges defined in ~\citep{yatawatta2015distributed}}
\label{fig:seq_distributed_SAGECal}
\end{figure}

A detailed understanding of these interactions is essential for the integration of SAGECal-MPI into the DALiuGE.
To port SAGECal-MPI to the DALiuGE framework, we made a deep investigation on the source code\footnote{git://git.code.sf.net/p/SAGECal/code} of the SAGECal-MPI. We obtain detailed data exchanges between the \textit{Master} and \textit{Slaves}, which are shown in Figure~\ref{fig:seq_distributed_SAGECal}. The feasibility of migrating SAGECal to the DALiuGE is also analyzed \citep{sdpmemo50}. 
Meanwhile, investigation is done on the accelerator support of execution framework \citep{sdpmemo77}.

In Figure~\ref{fig:seq_distributed_SAGECal}, the \textit{Master} process represents the ``Fusion center'' and several \textit{Slaves} processes represent the ``Agents'' as depicted in Figure \ref{fig:distributed_SAGECal}. The interactions between \textit{Master} and \textit{Slaves} are structured within a nested loop. The outer loop, shown as \textit{Timeslots Iteration}, iterates over data chunks, each of which consists of a fixed number \(tilesz\) of time samples. The inner loop contains \textit{ADMM Iterations}, the number, \(Nadmm\), of which is a predefined parameter. 

Within a particular ADMM iteration \(n\), the minimization of \(g'_i(J)\) is performed by a \textit{Slave} \(fi\) (i.e. inside the \texttt{sagefit\_vis}), which sends current the Lagrange multiplier \((Y_{fi})^n\) and the newly estimated Jones matrix \((J_{fi})^{n+1}\) to \textit{Master} upon completion. After receiving this information from all \textit{Slaves}, the \textit{Master} computes an updated global consensus \((Z)^{n+1}\), and broadcasts it back to all the \textit{Slaves}. The \textit{Slave} \(fi\) then updates its local Lagrange multiplier from \((Y_{fi})^{n}\) to \((Y_{fi})^{n+1}\) based on the ``goodness of fit'' of the smoothness constraint \([(J_{fi})^{n+1} - B_{fi}(Z)^{n+1}]\).

\section{Running SAGECal on the DALiuGE}

\label{sec:mpi}

% As analyzed in the previous sub-section that both \textit{Master} and \textit{Slaves} of the SAGECal are long running processes that explicitly maintain internal states, send/receive messages across multiple iterations, and manage data synchronization. 

% According to the requirements of the evaluation, we have been continuously optimizing SAGECal/DALiuGE so that it can fully utilize the characteristics of the DALiuGE and thus obtain fairer results.

Three different approaches have been developed for running SAGECal on the DALiuGE. 1) Coarse-grained porting by incorporating such ``stateful'' MPI applications directly into the DALiuGE execution framework. 2) Fine-grained porting using ``stateless'' Application and Data Drops. This decomposition will lead to a dataflow-style application that shifts the burden of state management completely to the DALiuGE. 3) Fine-grained porting with optimized Data Drops. This approach is basically the same as approach 2), but we have improved the Data Drop to make full use of memory for variable passing.

Regardless of the approach used, we ensure that the final processing result is consistent with SAGECal-MPI is the standard for porting and running the codes.

\subsection{Coarse-grained Porting: Drop Wrapping of MPI}\label{sec:mpiapp}

\begin{figure}[ht!]
\centering
\includegraphics[scale=0.27]{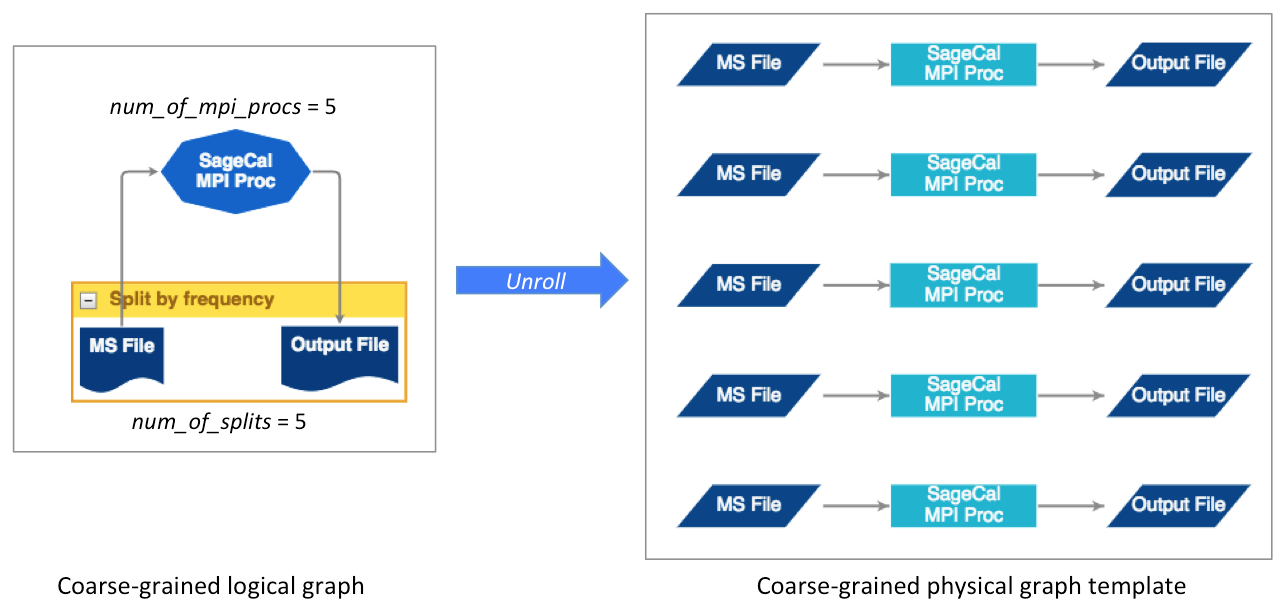}
\caption{Wrap the SAGECal-MPI using \texttt{MPIApp} Drops in the logical graph, which is then unrolled into the physical graph template with 5 MPI processes and split MeasurementSet files and output files.}
\label{fig:lg_mpi_drop}
\end{figure}

To quickly run MPI application on the DALiuGE, we developed a new Application Drop class, i.e., MPIApp Drop, to integrate any existing MPI application seamlessly. The basic principle of MPIApp Drop is quite simple. The octagon shape in the logical graph (Figure \ref{fig:lg_mpi_drop}) represents an \texttt{MPIApp}, which has a Drop property of \texttt{num\_of\_procs}, whose value is set to 5 in this example. During the graph translation stage \citep{wu2017daliuge}, five \texttt{MPIApp} Drops are generated. One of them represents the Master process, and the other four of them represent the Slave processes. Each \texttt{MPIApp} Drop has its own input and output files. For the Master, the input is the metadata file describing dataset locations, and the output is the final solution file. This option requires little modification or augmentation of the SAGECal-MPI  implementation for DALiuGE integration. In addition, during DALiuGE graph execution, the \texttt{MPIApp} Drop wrapper will monitor the execution progress of its enclosing MPI process, ensuring that each application is executed in the node assigned at scheduling time, and trigger events across edges of the graph to downstream Drops for cascaded execution. 

Obviously, such integration approach is a tricky porting approach. In this mode, DALiuGE is the equivalent of the MPI program launcher. The overall running efficiency is comparable to running MPI programs directly. However, in essence, this method of porting cannot give full play to the performance of DALiuge.

\subsection{Fine-grained Porting Based on File Data Drop}
\label{sec:graph}

In order to evaluate the DALiuGE more truthfully and fairly, we further rewrite the codes of SAGECal and make it to a real DALiuGE application.

The core of the DALiuGE is data-driven. Drop is a stateless program unit that starts running processing when Drop receives data and outputs data when processing is finished. 
Based on the detailed analysis of the SAGECal code, we disassembled the SAGECAL-MPI and constructed a series of fine-grained application Drops. 
Each of which represents either stateless computation (Application Drop) or persistent/transient datasets (Data Drop). These fine-grained component Drops were inherited from two standard Drops of the DALiuGE, i.e.,  \textit{BashShellApp} and \textit{FileDrop}.
\textit{BashShellApp} is used to revoke C/C++ programs and \textit{FileDrop} is used to transmit variables. 

\begin{figure}[ht!]
\centering
\includegraphics[scale=0.35]{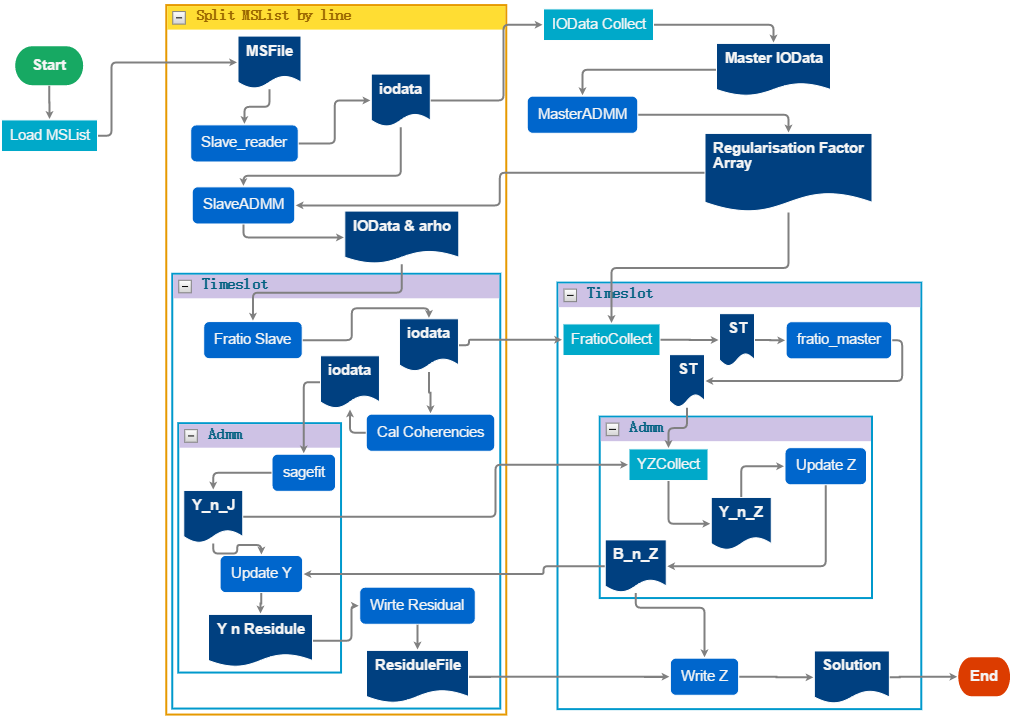}
\caption{Logical graph of the distributed SAGECal algorithm with BashShellApp and FileDrop.}
\label{fig:lg_dist_SAGECal_file}
\end{figure}

The logical graph of SAGECal in the DALiuGE (Figure \ref{fig:lg_dist_SAGECal_file}) includes around 30 Drops implemented with \textit{BashShellApp} and \textit{FileDrop}, which are the result of the decomposition of the original MPI-based distributed SAGECal application. \textit{BashShellApp} is provided as a wrapper of the actual executable program, and in each \textit{BashShellApp}, we need to define an entry for executable programs and related arguments. The general rule for creating these Drops is based on the communications between the Master and Slaves and between the application and the underlying I/O system (in this case, the file system). It should be noted that the implementation of such a decomposition requires considerable modifications to the existing MPI wrapper of the distributed SAGECal algorithm.
The logical graph of the distributed SAGECal in Figure \ref{fig:seq_distributed_SAGECal} is further translated into an executable physical graph.

The SAGECal-DALiuGE can use DALiuGE’s scheduling capability to run these physical graphs on distributed resources (rather than on a single node) in an optimal way in terms of total resource footprint, data movement volume, and execution time. However, the variable passing are based on file, which would significantly reduce the execution performance. 

\subsubsection{Fine-grained Porting Based on Variable Passing With Memory}

If variables passing can be performed between memories, it can obviously improve the performance. \texttt{DynlibApp}, which allows C/C++ programs to read and write variables in \texttt{MemoryDrop}, was developed. By using CTYPES, a foreign function library for Python, \texttt{DynlibApp} supports the manipulation of Input and Output DataDrops in C/C++ codes.

\begin{figure}[ht!]
\centering
\includegraphics[scale=0.38]{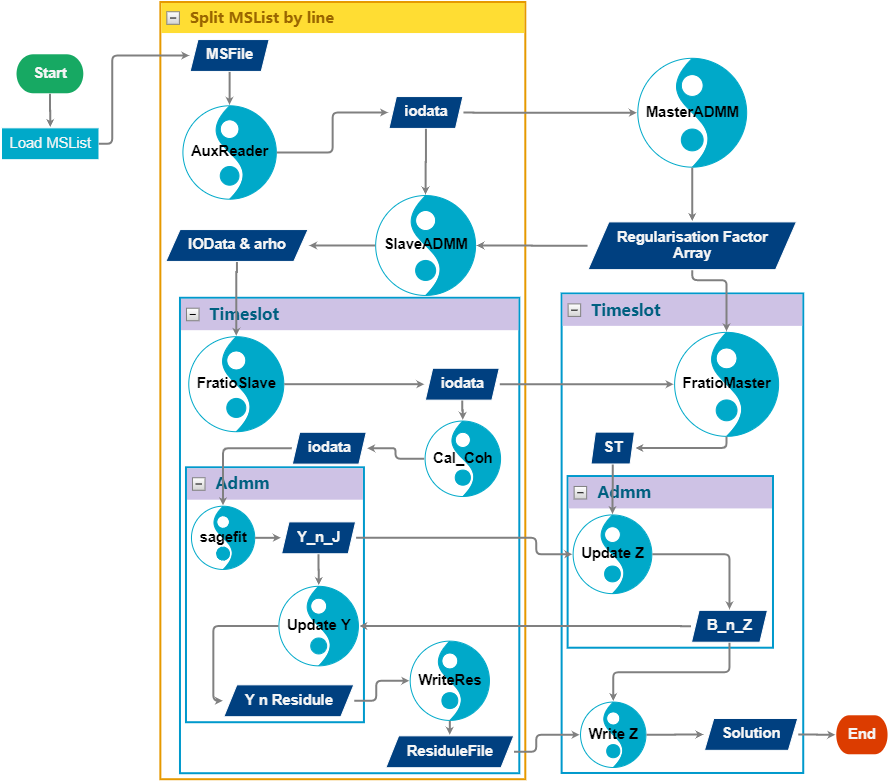}
\caption{Logical graph of the distributed SAGECal algorithm with DynlibApp and MemoryDrop}
\label{fig:lg_dist_SAGECal_mem}
\end{figure}

As shown in Figure \ref{fig:lg_dist_SAGECal_mem}, all the  \texttt{BashShellApp} in Figure \ref{fig:lg_dist_SAGECal_file} are replaced by \texttt{DynlibApp}. Meanwhile, we used \texttt{MemoryDrop} to replace all \texttt{FileDrop}. Each \texttt{DynlibApp} Drop needs to specify a dynamic link library file (``.so'' files in Linux). we create two classes, i.e., Master and Slave, and compile them into two ``.so'' files respectively, for invoking different functions. 

%  It means that a C/C++ program can invoke the methods (functions) of a Python MemoryDrop to receive and send the data. Thus, the difficulty of porting applications would be significantly decreased.
% According to the requirements of CTYPES, one property is considered to store the location of shared libraries for the \texttt{DynlibApp}.
% All the C/C++ programs should be compiled as ``.so'' files. In each program, two methods, \texttt{init()} and \texttt{run()}, must be implemented. When the Drop is executed, \texttt{init()} would be invoked first for the environment and variable initialization. The concrete algorithms implementation would be coded in \texttt{run()}. The two methods are declared as follows\footnote{ The detailed specifications are available in \textit{dlg\_app.h} at  https://github.com/ICRAR/daliuge}:

% \begin{lstlisting}[language=C, frame=single]
%     int init(dlg_app_info *app, const char ***params);
%     int run(dlg_app_info *app);
% \end{lstlisting}

\section{Performance Evaluation}

According to our experiments, the inputs and outputs of the DALiuGE Drops are exactly the same as the corresponding functions in MPI, which proofs a successful port with functional equivalence.
For performance comparison, we test SAGECal running on DALiuGE and MPI in the same environment with the same arguments.
Our experimental environment consists of four virtual machines on an internal computing cluster. Four virtual machines are installed with CentOS 7.3 operating system and have the same hardware configuration: Intel Xeon E5-2620 v2 CPUs, 2.1 GHz, 24 cores, 8 GB of memory, 60 GB of hard disk, and 1Gbps network bandwidth. In addition, another node is used to provide file-sharing service (via nfs-server) and web interface for submitting tasks.

\begin{figure}[ht!]
\centering
\includegraphics[scale=0.55]{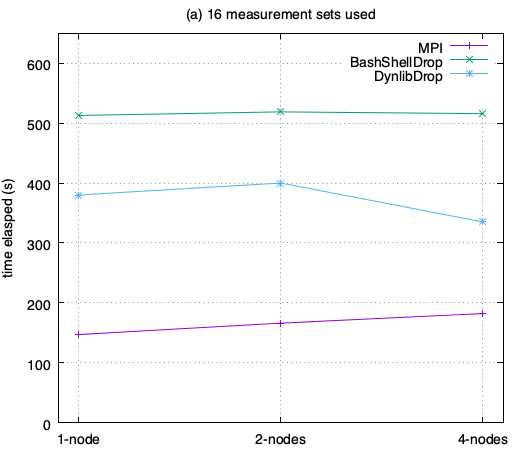}
\caption{Comparison of elapsed times for processing 16 measurement sets on varying nodes using three different methods (MPI, FileDrop, and MemoryDrop).}
\label{fig:eval_time1}
\end{figure}

\begin{figure}[ht!]
\centering
\includegraphics[scale=0.55]{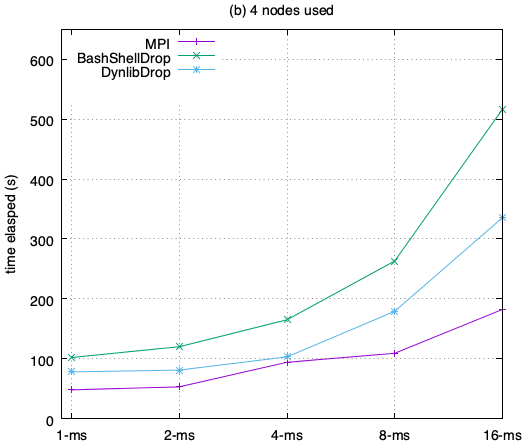}
\caption{Comparison of elapsed time for processing varying number of measurement sets using three different methods (MPI, FileDrop, and MemoryDrop).}
\label{fig:eval_time2}
\end{figure}

It can be seen through experiments that the SAGECal-MPI exhibits better performance than the DALiuGE version. Figure \ref{fig:eval_time1} shows our experimental results when dealing with 16 measurement-set files on 1,2,4 nodes. This is true when we port SAGECal to the DALiuGE.
In the case of a fixed number of measurement set files, as the number of processing nodes increases, the entire processing time of using the MPI version of SAGECal version does not decrease.
In the case of a fixed number of processing nodes, as the number of measurement set file increases, the entire processing time increases in three approaches, and the original MPI version performs best (see Figure \ref{fig:eval_time2}).

In order to avoid excessive transmission of variables, we output all variables of different Drops in the DALiuGE execution framework into different files stored in a shared directory. 
The subsequent Drops could open the specified file and obtain the variables. Nevertheless, this brings extra I/O time overhead to the execution process. We have tested eight measurement set files on two nodes. The total time used for I/O of shared files is about 16 seconds. The experiments shows that the time used for I/O of share files increases linearly with the number of measurement set files.

\section{Workload}

% \textcolor{red}{Comments:} There is very little discussion on the cost of the port in this
% paper, certainly too little to claim this a contribution. This would
% require at least a discussion on the man power required, as well as
% the cost in terms of performance loss. Such an analysis would likely
% be very valuable. The only discussion on this point is a remark on the
% ratio of code analysis versus writing code.

Four people have been involved in this work. They are all professional engineers of computer science with a little astronomy background. The leader of the group has extensive experience in radio interferometric data processing. To be more in line with the reality of the SKA software development, we did not invite SAGECal developers to participate in the work. We would also like to know whether the software engineers can understand the relevant radio interferometric algorithms and data flow when only documentations and codes are available.

We use the agile software development model to manage all of our work. According to their actual work, we count workload from three aspects such as algorithm analysis, coding and debugging (see Table \ref{table_workload}).

\begin{table}[htbp]
\centering
\label{table_workload}
\caption{Workloads in three porting cases. Unit=days. 1 Sprint = 2 weeks (10 work days). }
\begin{tabularx}{\linewidth}{p{1.5cm}|p{1.5cm}|p{1.3cm}|p{1.3cm}|X}
\toprule
\hline
\small
Approach No. & Algorithm \newline Analysis & Coding & Debug \newline and \newline Testing & Description \\ \midrule
1 & 1 & 4 & 2 & 2 task in a sprint. \\ 
2 & 12 & 4 & 6 & 6 tasks in 2 sprints  \\ 
3 & / & 6 & 5 & Followed the analysis results of Approach 2, 2 tasks in 1 sprint. \\
\bottomrule
\hline
\end{tabularx}
\end{table}

Obviously, Table~\ref{table_workload} answered the question we proposed in the section of introduction. The development on the DALiuGE is not too difficult and DROP wrapping can be done effectively after proper learning. In agile development the tasks scheduled in a Sprint are completed within the specified time. 

\section{Discussion}

Since the SDP consortium was established, it has been trying to establish a distributed computing system and build an execution framework for the SKA data processing. In terms of technology trends, 
using execution framework to process the super-massive data of the SKA is likely to be an inevitable choice. 
It is in this context that the DALiuGE was proposed and developed. 

However, according to our experiments, the DALiuGE execution framework has no advantage in operating performance. This result is also in line with the prior speculation. The advantages of the DALiuGE are fast task customization and a data-driven oriented processing model, which is probably more suitable for applications such as Science Regional Centre (SRC).

The port of the software was quite labor intensive. 
Actually, it is a strenuous work to port the MPI version of SAGECal to the DALiuGE. This work is not technically difficult, but the implementation consumes a lot of manpower. In fact, there is a huge gap between the DALiuGE and the mature parallel implementation such as MPI. 

The rationalization of experts from different disciplines for the final SKA software development is a matter of great concern. In this study, we did not ask any information from the author of the SAGECal. 
All algorithm analyses are dependant on the computer scientists themselves. To clarify the algorithm flow, divide and wrap the functions into the DALiuGE Drops, computer scientists in the study took a lot of work to analyze the principles and implementation of astronomical algorithms. They spent approxmate $55\%$ of full efforts in source code analysis, process analysis and data flow analysis, but only $20\%$ on porting and new codes design. In addition, DALiuGE's code debugging is inefficient because of no auxiliary tools.

\section{Conclusion}
\label{sec:conclusion}

We made a deep investigation on running a astronomical application on the DALiuGE execution framework so as to evaluate the fair performance of the DALiuGE. To achieve this goal, we ported a classical astronomical software, i.e., SAGECAL-MPI to the DALiuGE execution framework with three approaches. The study proposes some objective comments on the DALiuGE execution framework for future SKA software development, including the fact that the execution is not as efficient as MPI and the porting development is difficult. It is suggested that DALiuGE should find an adapted domain for further development.

All codes are released at \url{https://github.com/astroitlab/SAGECal-daliuge} and \url{https://github.com/astroitlab/SAGECal-daliuge-dynlib}.

\section{Acknowledgments}

This work is supported by the National SKA Program of China (2020SKA0110300), the National Natural Science Foundation of China (No.11903009, 11961141001), the Joint Research Fund in Astronomy (No. U1831204 and U1931141) under cooperative agreement between the National Natural Science Foundation of China (NSFC) and the Chinese Academy of Sciences (CAS), Yunnan Key Research and Development Program(2018IA054)

This work was also supported by resources provided by the Pawsey Supercomputing Centre with funding from the Australian Government and the Government of Western Australia. The authors wish to thank Mark Boulton for setting up the Pleiades MPI/SLURM environment for testing distributed SAGECal at ICRAR. 

The authors wish to thank Sarod Yatawatta for providing the SAGECal implementation and its associated test datasets. The authors wish to thank Ger van Diepen for providing information on SAGECal. The International Centre for Radio Astronomy Research (ICRAR) is a joint venture between Curtin University and The University of Western Australia with support and funding from the State Government of Western Australia.

% \section*{References}
\bibliographystyle{elsarticle-num}
\bibliography{mybibfile}

\end{document}